\documentclass{article}

\usepackage{arxiv}

\usepackage[utf8]{inputenc} 
\usepackage[T1]{fontenc}    
\usepackage{hyperref}       
\usepackage{url}            
\usepackage{booktabs}       
\usepackage{amsfonts}       
\usepackage{amssymb}        
\usepackage{amsmath}        
\usepackage{nicefrac}       
\usepackage{microtype}      
\usepackage{lipsum}		
\usepackage{graphicx}
\usepackage[numbers]{natbib}
\usepackage{doi}
\usepackage{wrapfig}

\title{Quantum algebra in R: the weyl package}

\author{ \href{https://orcid.org/0000-0001-5982-0415}{\includegraphics[width=0.03\textwidth]{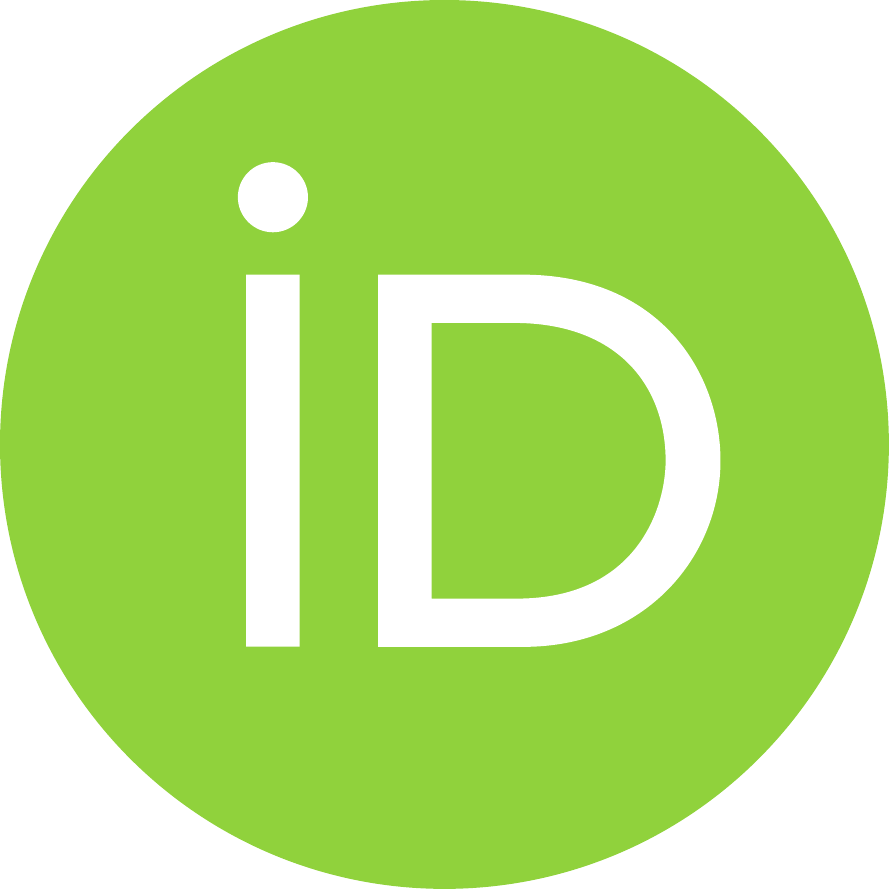}\hspace{1mm}Robin K. S.~Hankin}\thanks{\href{https://academics.aut.ac.nz/robin.hankin}{work};  
\href{https://www.youtube.com/watch?v=JzCX3FqDIOc&list=PL9_n3Tqzq9iWtgD8POJFdnVUCZ_zw6OiB&ab_channel=TrinTragulaGeneralRelativity}{play}} \\
 Auckland University of Technology\\
	\texttt{hankin.robin@gmail.com} \\
}

\renewcommand{\shorttitle}{Quantum algebra in R}

\hypersetup{
pdftitle={Quantum algebra in R},
pdfsubject={q-bio.NC, q-bio.QM},
pdfauthor={Robin K. S.~Hankin},
pdfkeywords={Quantum algebra, Weyl algebra, noncommutative algebra}
}

\usepackage{Sweave}
\begin{document}
\maketitle

\begin{abstract}
Weyl algebra is a simple noncommutative system used in quantum
mechanics.  Here I introduce the {\tt weyl} package, written in the R
computing language, which furnishes functionality for working with
univariate and multivariate Weyl algebras.  The package is available
on CRAN at \url{https://CRAN.R-project.org/package=weyl}.
\end{abstract}

\section{Introduction}

\setlength{\intextsep}{0pt}
\begin{wrapfigure}{r}{0.2\textwidth}
  \begin{center}
\includegraphics[width=1in]{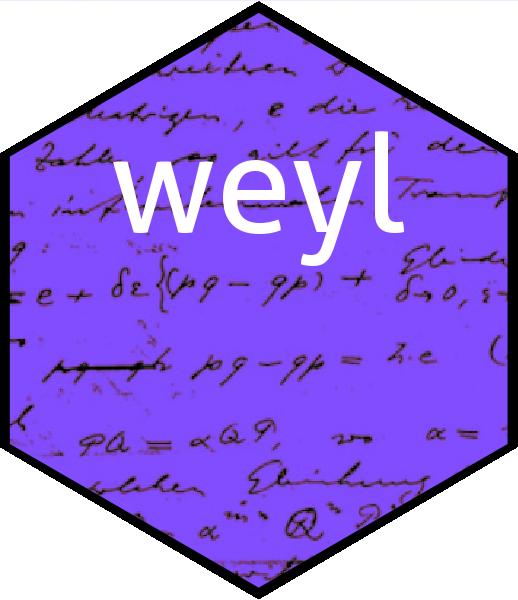}
  \end{center}
\end{wrapfigure}
Consider the vector space ${\mathcal A}$ of linear operators on
univariate functions; ${\mathcal A}$ can be made into an algebra,
conventionally called a Weyl algebra~\cite{coutinho1997}, where
multiplication (denoted by juxtaposition) is defined as operator
composition.  That is, given operators ${\mathcal O}_1,{\mathcal O}_2$
we define their product ${\mathcal O}_1{\mathcal O}_2$ by

\[({\mathcal O}_1{\mathcal O}_2)f={\mathcal O}_1({\mathcal O}_2f)\]

where $f$ is any univariate function.  Here we consider the algebra
generated by the set $\left\lbrace\partial,x\right\rbrace$ where
$\partial\colon f\longrightarrow f'$ [that is, $(\partial f)(x) =
f'(x)$] and $x\colon f\longrightarrow xf$ [that is, $(xf)(x) =
xf(x)$].  This is known as the (first) Weyl algebra.  We observe that
the Weyl algebra is not commutative: $\partial xf=(xf)'=f+xf'$ but
$x\partial f=xf'$, so $\partial x=x\partial+1$.  The algebra generated
by $\left\lbrace x,\partial\right\rbrace$ will include elements such
as $7\partial + 4\partial x\partial^3 x$, which would map $f$ to $7f'
+ 4\left(x\left(xf\right)'''\right)'$.  It can be shown that any
element of the Weyl algebra can be expressed in the standard form

\[
\sum_i a_i \partial^{p_i}x^{q_i}
\]

for real $a_i$ and nonnegative integers $p_i,q_i$.  Converting a
general word to standard form is not straightforward but we have

\[
\partial x^n = x^n\partial +  nx^{n-1}\]

and

\[
\partial^n x = x\partial^n +  n\partial^{n-1}.\]

We can apply these rules recursively to find standard form for
products $(\partial^i x^j)(\partial^l x^m)$.  Alternatively we may
follow Wolf~\cite{wolf1975} and use the fact that

\[
(\partial^i x^j)(\partial^lx^m)=
\sum_{r=0}^j{j\choose r}{l\choose r}\partial^{i+l-r}x^{j+m-r}.\]

These rules can be used to show, for example, that $7\partial +
4\partial x\partial^3 x$ can be expressed as $7\partial +
12x\partial^2 + 4x^2\partial^3$, which is in standard form.

\subsection{The package in use}

The above operators form part of the {\tt weyl} package, written in
the R computing language~\citep{rcore2022}, which is available on CRAN
at
\url{https://CRAN.R-project.org/package=weyl}~\cite{hankin2022_weyl}.
The package includes functionality to automate the above calculations
in an R-centric manner.  In particular, package idiom represents the
generating elements $\partial$ and $x$ of the first Weyl algebra as R
objects {\tt d} and {\tt x} respectively.  These may be manipulated
with standard arithemetic operations, and considering the example from
above we would have:

\begin{Schunk}
\begin{Sinput}
> library("weyl")
> 7*d + 4*x*d^3*x
\end{Sinput}
\begin{Soutput}
A member of the Weyl algebra:
 x d     val
 0 1  =    7
 1 2  =   12
 2 3  =    4
\end{Soutput}
\end{Schunk}

Above, the result is given in standard form.  We see the terms, one
per row, with coefficients in the rightmost column (viz $7,12,4$).
Thus the first row is $7\partial$, the second is $12x\partial^2$, and
the third is $4x^2\partial^3$.  We may choose to display the result in
symbolic form rather than matrix form:

\begin{Schunk}
\begin{Sinput}
> options(polyform=TRUE)
> 7*d + 4*x*d^3*x
\end{Sinput}
\begin{Soutput}
A member of the Weyl algebra:
+7*d +12*x*d^2 +4*x^2*d^3
\end{Soutput}
\end{Schunk}

which is arguably a more natural representation.  The package allows
one to use R semantics.  For example, consider $d_1=\partial x +
2\partial^3$ and $d_2=3+7\partial -5x^2\partial^2$.  Observing that
$d_1$ and $d_2$ are in standard form, package idiom to create these
operators would be:

\begin{Schunk}
\begin{Sinput}
> (d1 <- d*x + 2*d^3)
\end{Sinput}
\begin{Soutput}
A member of the Weyl algebra:
1 +x*d +2*d^3
\end{Soutput}
\begin{Sinput}
> (d2 <- 3 + 7*d  -5*x^2*d^2)
\end{Sinput}
\begin{Soutput}
A member of the Weyl algebra:
3 +7*d -5*x^2*d^2
\end{Soutput}
\end{Schunk}

(object {\tt d1} is converted to standard form automatically).  Observe
that, like the {\tt spray} package, the order of the terms is not defined.
We may apply the usual rules of arithmetic to these objects:

\begin{Schunk}
\begin{Sinput}
> d1*d2
\end{Sinput}
\begin{Soutput}
A member of the Weyl algebra:
3 +7*x*d^2 +3*x*d -15*x^2*d^2 -60*x*d^4 -5*x^3*d^3 +7*d +14*d^4 -54*d^3
-10*x^2*d^5
\end{Soutput}
\end{Schunk}

Standard R semantics operate, and it is possible to work with more
complicated expressions:

\begin{Schunk}
\begin{Sinput}
> options(polyform=TRUE)
> (d1^2 + d2) * (d2 - 3*d1)
\end{Sinput}
\begin{Soutput}
A member of the Weyl algebra:
-276*x*d^7 +28*d^7 +5*x^2*d^2 -732*d^6 -636*x*d^4 +28*d^4 -414*d^3
-63*x^2*d^3 +7*d -20*x^3*d^6 -24*d^9 -70*x*d^2 -20*x^2*d^8 +77*x^3*d^3
+20*x^4*d^4 -21*x*d +49*d^2 -198*x^2*d^5 +28*x*d^5
\end{Soutput}
\end{Schunk}

\subsection{Comparison with mathematica}

Mathematica can deal with operators and we may compare the two
systems' results for $\partial^2x\partial x^2$:

\begin{verbatim}
In[1] := D[D[x*D[x^2*f[x],x],x],x] // Expand

Out[1] := 4 f[x] + 14 x f'[x] + 8 x^2 f''[x] + x^3f'''[x]
\end{verbatim}

\begin{Schunk}
\begin{Sinput}
> x <- weyl(cbind(0,1))
> D <- weyl(cbind(1,0))
> x^2*D*x*D^2
\end{Sinput}
\begin{Soutput}
A member of the Weyl algebra:
4 +x^3*d^3 +8*x^2*d^2 +14*x*d
\end{Soutput}
\end{Schunk}

Above, we see agreement between {\tt weyl} and Mathematica, although the
terms are presented in a different order.

\section{Further Weyl algebras}

The package supports arbitrary multivariate Weyl algebras.  Consider:

\begin{Schunk}
\begin{Sinput}
> options(polyform=FALSE)    # revert to default print method
> set.seed(0)
> x <- rweyl()
> x
\end{Sinput}
\begin{Soutput}
A member of the Weyl algebra:
  x  y  z dx dy dz     val
  2  0  1  2  0  1  =    3
  0  1  2  2  0  1  =    2
  1  0  2  1  0  1  =    1
\end{Soutput}
\end{Schunk}

Above, object {\tt x} is a member of the operator algebra generated by
$\left\lbrace\partial_x,\partial_y,\partial_z,x,y,z\right\rbrace$.
Object {\tt x} might be expressed as $xz^2\partial_x\partial_z +
3x^2z\partial_x^2\partial_z + 2yz^2\partial_x^2\partial_z^2$ although
as ever the rows are presented in an implementation-dependent order.
We may verify associativity of multiplication:

\begin{Schunk}
\begin{Sinput}
> x <- rweyl(n=1,d=2)
> y <- rweyl(n=2,d=2)
> z <- rweyl(n=3,d=2)
> options(polyform=TRUE)
> x*(y*z)
\end{Sinput}
\begin{Soutput}
A member of the Weyl algebra:
+6*x*y^2*dx^2*dy +36*x*y^3*dx*dy^3 +2*x^2*y^4*dx*dy^4
+4*x^2*y^2*dx*dy^5 +36*x*y^2*dx*dy^2 +x*y^4*dx*dy^3 +4*x*y^3*dx*dy^2
+x^2*y^4*dx^2*dy^3 +12*x^2*y^2*dx*dy^2 +2*x^2*y^2*dx^2*dy^4
+2*x*y^2*dx*dy^4 +2*x*y^2*dx*dy +2*x^2*y^2*dx^2*dy +4*x^2*y^3*dx^2*dy^2
+3*x*y^4*dx^2*dy^3 +12*x^2*y^3*dx*dy^3 +6*x*y^4*dx*dy^4
+12*x*y^3*dx^2*dy^2
\end{Soutput}
\begin{Sinput}
> (x*y)*z
\end{Sinput}
\begin{Soutput}
A member of the Weyl algebra:
+2*x*y^2*dx*dy^4 +2*x^2*y^2*dx^2*dy^4 +3*x*y^4*dx^2*dy^3
+x^2*y^4*dx^2*dy^3 +12*x^2*y^2*dx*dy^2 +x*y^4*dx*dy^3
+2*x^2*y^4*dx*dy^4 +4*x^2*y^3*dx^2*dy^2 +2*x^2*y^2*dx^2*dy
+2*x*y^2*dx*dy +4*x*y^3*dx*dy^2 +12*x^2*y^3*dx*dy^3 +6*x*y^4*dx*dy^4
+12*x*y^3*dx^2*dy^2 +6*x*y^2*dx^2*dy +36*x*y^3*dx*dy^3
+36*x*y^2*dx*dy^2 +4*x^2*y^2*dx*dy^5
\end{Soutput}
\end{Schunk}

Comparing the two results above, we see that they apparently differ.
But the apparent difference is due to the fact that the terms appear
in a different order, a feature that is not algebraically meaningful.
We may verify that the expressions are indeed algebraically identical:

\begin{Schunk}
\begin{Sinput}
> x*(y*z) - (x*y)*z
\end{Sinput}
\begin{Soutput}
A member of the Weyl algebra:
the NULL multinomial of arity 4
\end{Soutput}
\begin{Sinput}
> options(polyform=FALSE)  # revert to default print method
\end{Sinput}
\end{Schunk}

The package can deal with arbitrarily high dimensional Weyl algebras.
For exmaple:

\begin{Schunk}
\begin{Sinput}
> (x9 <- rweyl(dim=9))
\end{Sinput}
\begin{Soutput}
A member of the Weyl algebra:
  1  2  3  4  5  6  7  8  9 d1 d2 d3 d4 d5 d6 d7 d8 d9     val
  1  2  2  1  0  1  2  2  2  0  2  0  0  0  0  2  1  1  =    3
  2  0  1  1  1  1  0  1  1  0  0  2  1  0  1  2  1  2  =    2
  0  0  1  2  1  1  1  2  2  2  0  1  0  0  2  0  2  1  =    1
\end{Soutput}
\end{Schunk}

Above we see a member of the ninth Weyl algebra; see how the column
headings no longer use the {\tt x y z} notation and revert to numeric
labels.  Symbolic notation is available but can be difficult to read:

\begin{Schunk}
\begin{Sinput}
> options(polyform=TRUE)
> x9
\end{Sinput}
\begin{Soutput}
A member of the Weyl algebra:
+3*x1*x2^2*x3^2*x4*x6*x7^2*x8^2*x9^2*d2^2*d7^2*d8*d9
+2*x1^2*x3*x4*x5*x6*x8*x9*d3^2*d4*d6*d7^2*d8*d9^2
+x3*x4^2*x5*x6*x7*x8^2*x9^2*d1^2*d3*d6^2*d8^2*d9
\end{Soutput}
\begin{Sinput}
> options(polyform=FALSE)  # revert to default print method
\end{Sinput}
\end{Schunk}

\section{Derivations}

A {\em derivation} $D$ of an algebra ${\mathcal A}$ is a linear operator
that satisfies $D(d_1d_2)=d_1D(d_2) + D(d_1)d_2$, for every
$d_1,d_2\in{\mathcal A}$.  If a derivation is of the form $D(d) =
[d,f] = df-fd$ for some fixed $f\in{\mathcal A}$, we say that $D$ is
an {\em inner} derivation:

\[
D(d_1d_2) =
d_1d_2f-fd_1d_2 =
d_1d_2f-d_1fd_2 + d_1fd_2-fd_1d_2 = 
d_1(d_2f-fd_2) + (d_1f-fd_1)d_2 =
d_1D(d_2) + D(d_1)d_2
\]

Dirac showed that all derivations are inner derivations for some
$f\in{\mathcal A}$.  The package supports derivations:

\begin{Schunk}
\begin{Sinput}
> f <- rweyl()
> D <- as.der(f)  # D(x) = xf-fx
\end{Sinput}
\end{Schunk}

Then

\begin{Schunk}
\begin{Sinput}
> d1 <- rweyl()
> d2 <- rweyl()
> D(d1*d2) == d1*D(d2) + D(d1)*d2
\end{Sinput}
\begin{Soutput}
[1] TRUE
\end{Soutput}
\end{Schunk}

\section{Low-level considerations and generalizations}

In the package, the product is customisable.  In general, product {\tt
a*b} [where {\tt a} and {\tt b} are {\tt weyl} objects] is dispatched
to the following sequence of functions:

\begin{itemize}
\item {\tt weyl\_prod\_multivariate\_nrow\_allcolumns()}
\item {\tt weyl\_prod\_multivariate\_onerow\_allcolumns()}
\item {\tt weyl\_prod\_multivariate\_onerow\_singlecolumn()}
\item {\tt weyl\_prod\_univariate\_onerow()}
\item {\tt weyl\_prod\_helper3()} (default)
\end{itemize}

In the above, ``univariate" means "generated by $\left\lbrace
x,\partial_x\right\rbrace$" [so the corresponding {\tt spray} object
has {\em two} columns]; and ``multivariate" means that the algebra is
generated by more than one variable, typically something like
$\left\lbrace x,y,z,\partial_x,\partial_y,\partial_z\right\rbrace$.

The penultimate function {\tt weyl\_prod\_univariate\_onerow()} is
sensitive to option {\tt prodfunc()} which specifies the recurrence
relation used.  This defaults to {\tt weyl\_prod\_helper3()}:

\begin{Schunk}
\begin{Sinput}
> weyl_prod_helper3
\end{Sinput}
\begin{Soutput}
function (a, b, c, d) 
{
    f <- function(r) {
        factorial(r) * choose(b, r) * choose(c, r)
    }
    ind <- numeric(0)
    val <- numeric(0)
    for (r in 0:b) {
        ind <- rbind(ind, c(a + c - r, b + d - r))
        val <- c(val, f(r))
    }
    spray(ind, val, addrepeats = TRUE)
}
<bytecode: 0x558b05c73dc8>
<environment: namespace:weyl>
\end{Soutput}
\end{Schunk}

Function {\tt weyl\_prod\_helper3()} follows Wolf.  This gives the
univariate concatenation product $(\partial^a x^b)(\partial^c x^d)$ in
terms of standard generators:

\[
\partial^a x^b \partial^c x^d=\sum_{r=0}^b
r!{b\choose r}{c\choose r}
\partial^{a+c-r}x^{b+d-r}
\]

The package also includes lower-level function {\tt weyl\_prod\_helper1()}
implementing $\partial^a x^b \partial^c
x^d=\partial^ax^{b-1}\partial^cx^{d+1} +
c\partial^ax^{b-1}\partial^{c-1}x^d$ (together with suitable
bottoming-out).  I expected function {\tt weyl\_prod\_helper3()} to be much
faster than {\tt weyl\_prod\_helper1()} but there doesn't seem to be much
difference between the two.

\section{Generalized commutator relations}

We can exploit this package customisability by considering, instead of
$\left\lbrace x,\partial\right\rbrace$, the algebra generated by
$\left\lbrace e,\partial\right\rbrace$, where $e$ maps $f$ to $e^xf$:
if $f$ maps $x$ to $f(x)$, then $ef$ maps $x$ to $e^xf(x)$.  We see
that $\partial e-e\partial=e$.  With this, we can prove that
$\partial^ne=e(1+\partial)^n$ and $e^n\partial=e^n\partial+ne^n$ and,
thus

\[
(e^a\partial^b)(e^c\partial^d)
=e^{a+1}(1+\partial)^be^{c-1}\partial^d
=e^{a}\partial^{b-1}e^{c}\partial^{d+1}+ce^{a}\partial^{b-1}e^{c}\partial^d
\]

We may implement this set in package idiom as follows:

\begin{Schunk}
\begin{Sinput}
> `weyl_e_prod` <- function(a,b,c,d){
+     if(c==0){return(spray(cbind(a,b+d)))}
+     if(b==0){return(spray(cbind(a+c,d)))}
+     return(
+     Recall(a,b-1,c,d+1) +
+     c*Recall(a,b-1,c,d)  # cf: c*Recall(a,b-1,c-1,d)) for regular Weyl algebra
+ ) }
\end{Sinput}
\end{Schunk}

Then, for example, to calculate $\partial^2e=e(1+2\partial+\partial^2)$:

\begin{Schunk}
\begin{Sinput}
> options(prodfunc = weyl_e_prod) 
> options(weylvars = "e")  # changes print method
> d <- weyl(spray(cbind(0,1)))
> e <- weyl(spray(cbind(1,0)))
> d*d*e
\end{Sinput}
\begin{Soutput}
A member of the Weyl algebra:
 e d     val
 1 0  =    1
 1 1  =    2
 1 2  =    1
\end{Soutput}
\begin{Sinput}
> d^2*e
\end{Sinput}
\begin{Soutput}
A member of the Weyl algebra:
 e d     val
 1 0  =    1
 1 1  =    2
 1 2  =    1
\end{Soutput}
\end{Schunk}

By way of verification:

\begin{Schunk}
\begin{Sinput}
> d^5*e == e*(1+d)^5
\end{Sinput}
\begin{Soutput}
[1] TRUE
\end{Soutput}
\end{Schunk}

which verifies that indeed $\partial^5e=e(1+\partial)^5$.  Another
verification would be to cross-check with Mathematica, here working with
$\partial e\partial^2e$:

\begin{verbatim}
In[1] := D[Exp[x]*D[D[Exp[x]*f[x],x],x],x]

Out[1] := 2E^2x f[x] + 5E^2x f'[x] + 4E^2xf''[x] + E^2x f'''[x]
\end{verbatim}

\begin{Schunk}
\begin{Sinput}
> options(polyform = TRUE)
> d*e*d^2*e
\end{Sinput}
\begin{Soutput}
A member of the Weyl algebra:
+2*e^2 +5*e^2*d +4*e^2*d^2 +e^2*d^3
\end{Soutput}
\end{Schunk}

We can manipulate more complicated expressions too.  Suppose we want to
evaluate $(1+e^2\partial)(1-5e^3\partial^3)$:

\begin{Schunk}
\begin{Sinput}
> o1 <- weyl(spray(cbind(2,1)))
> o2 <- weyl(spray(cbind(3,3)))
> options(polyform = FALSE)
> (1+o1)*(1-5*o2)
\end{Sinput}
\begin{Soutput}
A member of the Weyl algebra:
 e d     val
 5 3  =  -15
 3 3  =   -5
 5 4  =   -5
 2 1  =    1
 0 0  =    1
\end{Soutput}
\end{Schunk}

And of course we can display the result in symbolic form:

\begin{Schunk}
\begin{Sinput}
> options(polyform = TRUE)
> (1+o1)*(1-5*o2)
\end{Sinput}
\begin{Soutput}
A member of the Weyl algebra:
1 -15*e^5*d^3 -5*e^3*d^3 -5*e^5*d^4 +e^2*d
\end{Soutput}
\end{Schunk}

\begin{Schunk}
\begin{Sinput}
> options(polyform = NULL) # restore default print method
\end{Sinput}
\end{Schunk}

\section{Computational implementation and notes on {\tt disordR}
discipline}

The package stores {\tt weyl} objects using the {\tt
spray}~\cite{hankin2022_spray_arxiv} class for sparse arrays.
Addition is inherited from {\tt spray}; multiplication is specific to
the {\tt weyl} package.  Thus the coefficients of a {\tt weyl} object,
and the rows of its index matrix, are stored in an
implementation-specific order.  Extraction and replacement use {\tt
disordR} discipline~\cite{hankin2022_disordR_arxiv}.  A short example
follows in the context of the {\tt weyl} package; much more extensive
and detailed discussions are given by
Hankin~\cite{hankin2022_disordR_arxiv,hankin2022_mvp_arxiv}.

\subsection{An illustrative session}

Here I show how {\tt disordR} discipline is used in a typical R
session.  First we create a moderately complicated {\tt weyl} object:

\begin{Schunk}
\begin{Sinput}
> options(weylvars = NULL)   # revert to default names
> (W <- weyl(spray(matrix(c(0,1,1,1,1,2,1,0),2,4),2:3))^2)
\end{Sinput}
\begin{Soutput}
A member of the Weyl algebra:
  x  y dx dy     val
  2  2  4  0  =    9
  2  2  3  0  =   18
  0  2  2  2  =    4
  2  2  2  0  =    9
  1  2  3  1  =   12
  1  2  2  0  =    6
  1  2  3  0  =    6
  1  2  2  1  =    6
  0  2  2  1  =    4
\end{Soutput}
\end{Schunk}

The coefficients of {\tt W} may be extracted:

\begin{Schunk}
\begin{Sinput}
> coeffs(W)
\end{Sinput}
\begin{Soutput}
A disord object with hash ef3b76da15a19ac6dd3ba83e2ec6b436a0f975f6 and elements
[1]  9 18  4  9 12  6  6  6  4
(in some order)
\end{Soutput}
\end{Schunk}

The object returned is a {\tt disord} object.  There is no way to extract
(e.g.) the first coefficient, for the order of the matrix rows is not
defined.  If we try we will get an error:

\begin{Schunk}
\begin{Sinput}
> coeffs(W)[1]
\end{Sinput}
\begin{Soutput}
Error in .local(x, i, j = j, ..., drop) : 
  if using a regular index to extract, must extract each element once and once only (or none of them)
\end{Soutput}
\end{Schunk}

However, questions such as ``give all coefficients greater than 6" are
perfectly well defined:

\begin{Schunk}
\begin{Sinput}
> o <- coeffs(W)
> o[o>6]
\end{Sinput}
\begin{Soutput}
A disord object with hash ebf6f5624c14b9166f702b460f5f202351ed5026 and elements
[1]  9 18  9 12
(in some order)
\end{Soutput}
\end{Schunk}

Extraction works as expected.  Using recent improvements in the
{\tt disordR} package, we take all coefficients less than 7 and add 100 to
them:

\begin{Schunk}
\begin{Sinput}
> coeffs(W)[coeffs(W)<7] <- coeffs(W)[coeffs(W)<7] + 100
> W
\end{Sinput}
\begin{Soutput}
A member of the Weyl algebra:
  x  y dx dy     val
  0  2  2  1  =  104
  1  2  2  1  =  106
  1  2  3  0  =  106
  1  2  2  0  =  106
  1  2  3  1  =   12
  2  2  2  0  =    9
  0  2  2  2  =  104
  2  2  3  0  =   18
  2  2  4  0  =    9
\end{Soutput}
\end{Schunk}

\section{Conclusions and further work}

The {\tt weyl} package implements Weyl algebra using real
coefficients.  However, in quantum mechanics one typically works with
the operator $p_j$, defined as $-i\partial_j$ (where $i$ is the
imaginary unit of the complex plane, $i^2=-1$); thus
$x_jp_j-p_jx_j=-i$.  Further work might include admitting complex
coefficients to accommodate this parametrization.

\bibliographystyle{apalike}
\bibliography{weyl_arxiv}

\end{document}